\documentclass[superscriptaddress,groupedaddress,nofootnoteinbib,12pt,english]{article}
\usepackage{jcappub}
\usepackage{graphicx}
\usepackage{bm}        
\usepackage{amssymb}   
\usepackage{hyphenat}
\usepackage{mathdots}

\usepackage{ textcomp }
\input epsf

\def\be{\begin{equation}}
\def\ee{\end{equation}}
\def\ba{\begin{eqnarray}}
\def\ea{\end{eqnarray}}
\def\nn{\nonumber}
\def\K{\mathcal{K}}
\def\ie{{\it i.e. }}
\def\({\left(}
\def\){\right)}
\def\O{\mathcal{O}}
\def\mpl{M_{\rm Pl}}

\def\p{{\cal P}}

\def\L*{{\cal L}_*}
\def\L{\mathcal{L}}
\def\({\left(}
\def\){\right)}
\def\ie{{\it i.e. }}
\def\nn{\nonumber}
\def\p{\partial}

\def\p{\partial}

\def\<{\langle}
\def\>{\rangle}
\usepackage{amsmath}
\usepackage{amsfonts}
\usepackage{verbatim}
\usepackage{setspace}
\begin{document}
\thispagestyle{empty}

\title{Superluminality in the Bi- and Multi- Galileon}
\author{Paul de Fromont${}^{a}$,  Claudia de Rham${}^{b}$, Lavinia Heisenberg${}^{b,c}$ and Andrew Matas${}^{b}$}


\affiliation{${}^{a}$Ecole Normale Sup\'erieure de Lyon, Universit\'e Claude Bernard Lyon I, Lyon, France\\
${}^b$Department of Physics, Case Western Reserve University, 10900 Euclid Ave, Cleveland, OH
44106, USA\\
${}^{c}$D\'epartement de Physique Th\'eorique and Center for Astroparticle Physics,
 Universit\'e de Gen\`eve, 24 Quai E. Ansermet, CH-1211,  Gen\`eve, Switzerland
}

\abstract{We re-explore the Bi- and Multi-Galileon models with trivial asymptotic conditions at infinity and show that propagation of superluminal fluctuations is a common and unavoidable feature of these theories, unlike previously claimed in the literature.
We show that all Multi-Galileon theories containing a Cubic Galileon term exhibit superluminalities at large distances from a point source, and that even if the Cubic Galileon is not present one can always find sensible matter distributions in which there are superluminal modes at large distances. In the Bi-Galileon case we explicitly show that there are always superluminal modes around a point source even if the Cubic Galileon is not present. Finally, we briefly comment on the possibility of avoiding superluminalities by modifying the asymptotic conditions at infinity.}

\maketitle

\setcounter{footnote}{0}

\newpage

\section{Introduction}

The combination of independent observational sources such as the CMB, supernovae, lensing and baryon acoustic oscillations indicate that the Universe is expanding at an increasing rate, possibly driven by dark energy, \cite{Hinshaw:2012fq}. Despite a significant theoretical and observational efforts the physical origin of the accelerated expansion is a mystery. Not only its discovery but also the resolution to the puzzle of its origin are an unexaggerated merit of Nobel-prizes. Explanatory attempts fall into three primary categories.

The first solution consists of considering a small cosmological constant $\Lambda$ with a constant energy density giving rise to an effective repulsive force between cosmological objects at large distances \cite{Peebles:2002gy}. If we assume that the cosmological constant correspond to the vacuum energy density, then the theoretical expectations for the vacuum energy density caused by fluctuating quantum fields, differs from the observational bounds on $\Lambda$ by up to 120 orders of magnitude. This giant mismatch between the theoretically computed high energy density of the vacuum and the low observed value remains for more almost a century one of the most challenging puzzles in physics and is called the cosmological constant problem.

The second solution could for instance consist in introducing new dynamical degrees of freedom by invoking new fluids $T_{\mu\nu}$ with negative pressure. Quintessence is one of the important representatives of this class of modification. The acceleration is due to a scalar field whose kinetic energy is small in comparison to its potential energy, causing dynamical equation of state with initially negative values \cite{Doran:2000jt}. This class of theories might as well as the cosmological constant exhibit fine-tuning problems.

Alternatively, the third solution would correspond to explaining the acceleration of the Universe by changing the geometrical part of Einstein's equations. In particular, weakening gravity on cosmological scales might not only be responsible for a late-time speed-up of the Hubble expansion but could also tackle the cosmological constant problem. Such scenarios could arise in massive gravity or in higher-dimensional frameworks.

Concerning theories of higher dimensional theories, the Dvali-Gabadadze-Porrati (DGP) model has set the stage for large scale modified theories of gravity \cite{DGP}. In this framework our Universe is a three dimensional brane embedded in a five-dimensional bulk. In this higher dimensional setup, the effective four-dimensional graviton on the brane carries more degrees of freedom, amongst them the scalar degree of freedom carrying most of the physical impact of this changement. This is explained by the graviton acquiring a soft mass $m$ which limits its effective range whereas on small scales one recovers General Relativity (GR) through the so called Vainshtein mechanism: the additional degree of freedom, the helicity-0 mode, is decoupled from the gravitational dynamics via nonlinear interactions of the helicity-0 mode itself, \cite{Vainshtein:1972sx}.  This decoupling of the nonlinear helicity-0 mode is manifest in the limit where the four and five dimensional Plank scales are sent to infinity and the soft graviton mass $m\to 0$ while the strong coupling scale $\Lambda = (\mpl m^2)^{1/3}$ is kept fixed, enabling us to treat the usual helicity-2 mode of gravity linearly while the helicity-0 mode $\pi$ requires non-linear treatment, \cite{Luty:2003vm}. The achievement of the DGP model was the existence of a self-accelerating solution sourced by the graviton own degree of freedom, the helicity-0 mode.  As promising as this was, it was realized soon that this branch of solution was plagued by ghost-like instabilities \cite{DDG,Koyama:2005tx,Charmousis:2006pn}. This issue could be avoided for instance including Gauss-Bonnet terms in the bulk, \cite{deRham:2006pe}.

  Galilean invariant interactions were then introduced to extend the decoupling limit of DGP-gravity \cite{Nicolis:2008in}. This Galileon model relies strongly on the symmetry of the helicity-0 mode $\pi$, which is invariant under internal galileon- and shifting transformations $\pi\to\pi+b_\mu x^\mu+c$. This symmetry can be regarded as residuals of the 5-dimensional Poincar\'e invariance in induced gravity braneworld models. These symmetries, along with the postulate of the absence of ghosts, imposes drastic restriction on the allowed effective Galileon Lagrangians, as there is a total number of only five derivative interactions fulfilling these conditions in four dimensions (omitting the tadpole). In the context of 5-dimensional braneworld models, these Galilean invariant derivative interactions appear as consequences of Lovelock invariants \cite{deRham:2010eu}. For a review on the Galileon see Ref.~\cite{deRham:2012az}.
 The interesting feature of Galileons to allow self-accelerating de Sitter solutions while providing a ghost-free theory \cite{Nicolis:2008in,Silva:2009km} has generated a flurry of investigations, targeted at cosmological models and their observational signatures \cite{Khoury:2009tk}, inflationary models \cite{Creminelli:2010ba,Burrage:2010cu,Mizuno:2010ag,Hinterbichler:2011qk}, gravitational lensing \cite{Wyman:2011mp}, spherically symmetric solutions in the vicinity of compact sources \cite{Berezhiani:2013dw}, Binary Pulsars \cite{deRham:2012fw,deRham:2012fg}, etc\ldots.

 While cosmological models based on Galileon-gravity were commonly restricted to spatially flat backgrounds, there are also possibilities of introducing generalisations for non-flat models. One approach being a direct covariantization of the decoupling limit \cite{Chow:2009fm,deRham:2011by}. Naive covariantisation, however, can give rise to ghost-like terms in the equation of motion, which can be remedied by a unique non-minimal coupling between $\pi$ and the curvature, \cite{Deffayet:2009wt,Deffayet:2009mn,Deffayet:2011gz}. The explicit formalism was derived in Refs.~\cite{Deffayet:2009wt,Deffayet:2009mn} and the resulting covariantized Galileons are also consistent with a higher-dimensional setup \cite{deRham:2010eu}.
Generalizations to  maximally symmetric backgrounds with a new Galileon symmetry have been constructed successfully in Refs.~\cite{Goon:2011qf}.

Another interesting point to mention is that Galileon-type interaction terms naturally arise in theories of massive gravity, which has, in addition, been constructed to be ghost-free be it in three dimensions, \cite{deRham:2011ca} or for an generalized Fierz-Pauli action in four dimensions \cite{deRham:2010kj,deRham:2010ik}.

Finally, another important property of the Galileon interactions is the fact that they are not renormalizable. In other words, the Galileon coupling constants can be technically natural tuned to any value and remain stable under quantum corrections \cite{Luty:2003vm,Nicolis:2004qq,Hinterbichler:2010xn,deRham:2012ew}.

Despite the fact that the Galileon exhibits a broad and interesting phenomenology, they witness a potentially worrying phenomenon, namely the fluctuations of the Galileon field can propagate superluminally in the regime of interest \cite{Hinterbichler:2009kq,Nicolis:2008in,Goon:2010xh}. Since the Galileon was introduced as an extention of the decoupling limit of DGP-gravity, the DGP model shares the same phenomenon of superluminal fluctuations \cite{Adams:2006sv}. Superluminal modes are not only generic to Galileon \cite{Nicolis:2009qm} but also to massive gravity \cite{Dubovsky:2005xd,Gruzinov:2011sq,deRham:2011pt,Deser:2012qx}.

Theories with superluminal fluctuations are sick if they also allow for acausality, and configurations with Close-Timelike-Curves (CTCs) are present. Nevertheless there are cases in which the superluminal fluctuations come with their own metric and causal structure, which can be very different to that felt by photons, and the causal cones of these fluctuations might even lie outside the causal cones of photons. Regardless of all this, the causal structure of the spacetime can be protected \cite{Babichev:2007dw} if there exists one foliation of spacetime into surfaces which can be considered as Cauchy surfaces for both metrics.
In theories of  Galilean invariant interactions it is possible to construct CTCs within the naive regime of validity of the effective field theory (as is also the case in GR). Nevertheless, as it has been shown in \cite{Burrage:2011cr}, the CTCs never arise since the Galileon inevitably becomes infinitely strongly coupled implying an infinite amount of backreaction. The backreaction on the background for the Galileon field breaks down the effective field theory and forbids the formation of the CTC through the backreaction on the spacetime geometry. The setup of background solutions with CTCs becomes unstable with an arbitrarily fast decay time. As a result, theories of  Galilean invariant interactions satisfy a direct analogue of Hawking's chronology protection conjecture, see Ref.~\cite{Burrage:2011cr}.

The single Galileon scalar field theory has been generalized to a multitude of interacting Galileon fields whose origin again can be traced back to Lovelock invariants in the higher co-dimension bulk, \cite{Hinterbichler:2010xn}, or such as in Cascading Gravity, \cite{deRham:2007xp,Padilla:2010de,Padilla:2010tj}. Furthermore, they have been extended to arbitrary even p-forms whose field equations still only contain second derivatives \cite{Deffayet:2010zh}.

In this paper, we scrutinize the superluminality in Multi-Galileon theories and argue that in these models, the existence of the Vainshtein mechanism about a static spherically symmetric source comes hand in hand with the existence of superluminal modes. The argument goes as follows:
\begin{enumerate}
\item{\bf Superluminalities from the Cubic Galileon for a localized point-source:} We first show that the mere presence of Cubic Galileon interactions guaranties the superluminal propagation of modes in either the radial or the orthoradial direction far away from a point source. \\
    This is intrinsic to the fact that for such configurations, at least one field falls as $1/r$ at large distance as expected from the Coulomb potential. For that behaviour, the matrix encoding the temporal perturbations vanishes at infinity  while the orthoradial and radial perturbations arise with opposite sign. This property is independent of the number of Galileon fields present.
\item{\bf Superluminality from the Quartic Galileon for extended static spherically symmetric sources:} Since the previous result is ubiquitous to any Cubic Galileon interactions,  the only  possible way to avoid superluminalities at large distances, is to set all the Cubic Galileon interactions to zero. In that case, we show that the Quartic Galileon always lead to some superluminalities at large distances in either the radial or the orthoradial direction again  when considering a sensible extended source.
\item{\bf Superluminality from the Quartic Bi-Galileon for a localized point-source:} Even if we restrict ourselves to point sources, we show that the Quartic Galileon always lead to the propagation of at least one  superluminal radial mode for some range of $r$. This result contradicts previous claims found in the litterature.

 Our result relies crucially on the assumption that 1. the field decay as the Coulomb potential at infinity, 2. that no ghost are present  and 3. that the Vainshtein mechanism is active (\ie the Quartic Galileon interactions dominate over quadratic kinetic terms near the source). The derivation of our generic result relies on the interplay between the behaviour of the field at large and at small distances.
\item{\bf Superluminality from the Cubic Bi-Galileon for a localized point-source:} Finally, we show that near a localized source superluminalities are also present in the radial direction in a theory which includes only the Cubic Galileon.
\end{enumerate}
The rest of the paper is organized as follows: We start in section \ref{BiGal} with a summary of the formalism of the Bi- and Multi-Galileon. We then present the analysis needed for the study of the propagation of fluctuations around spherical symmetric backgrounds in section \ref{pert}.
In section \ref{large-distance} we study the perturbations around the background generated by a point mass at large distances from that source.  We show that there is always one mode which propagates superluminaly whenever the Cubic Galileon is present, regardless of the number of Galileons present in the theory. We also find that there are sensible source distributions around which there will always be a superluminal mode at large distances even if the Cubic Galileon is absent, for any number of Galileons. In section \ref{QuarticGalileonsection} we then consider more closely the case of a point mass source when the Cubic Galileon is absent. In particular we study the short distance behaviour around a point mass background in the Bi-Galileon theory, and we find that there is again always a superluminal mode. In the case of vanishing asymptotic conditions $\pi \to 0$ the existence of the Vainshtein mechanism comes hand in hand with the existence of superluminal modes. This constitutes a No-go theorem showing that superluminal modes are generic to Galileon theories. Finally in the discussion section, we comment on the only known loophole to this argument, which is to have non-vanishing asymptotic conditions for the field at infinity. This kind of set up can arise naturally in theories such as massive gravity which reduce to the Galileon theory in some limit.

In the rest of this work, we adopt the same notation as in \cite{Padilla:2010de} and demonstrate in this language that superluminalities can never be avoided\footnote{So long as the Galileon interactions dominate near the source (\ie, as long as the model exhibits a viable Vainshtein mechanism) and as long as one considers trivial asymptotic conditions at infinity.} in a consistent Galileon model.

\section{The Bi- and Multi-Galileon model}\label{BiGal}
In this work we consider the most general Multi-Galileon theory, in four dimensions,  which is conformally coupled to matter. This model  consists of $N$ coupled scalar fields, $\pi_1,\dots,\pi_N$.
For simplicity we neglect gravity in our analysis and study the theory on Minkowski space-time. Similarly to Galileon theories \cite{Nicolis:2008in}, the Multi-Galileon theory is invariant under internal Galilean and shift transformations $\pi_1\to\pi_1+b_1^\mu x_\mu +c_1,\;\cdots\;, \pi_N\to\pi_N+b_N^\mu x^\mu +c_N$. Without loss of generality, we couple only one of the N scalar fields to the trace of the stress energy tensor\footnote{One can always rotate the field space $\pi_1, \cdots, \pi_N$ so as to couple only one field to the trace of the stress-energy tensor. This would not be possible if more subtle coupling to matter were present, for instance $\p_\mu \pi \p_\nu \pi T^{\mu\nu}$ as is the case in Massive Gravity \cite{deRham:2010ik}. However this coupling cancels at the background level for static spherically symmetric sources.},
\begin{equation}
\mathcal L= \L_2+\L_3+\L_4+\L_5+\pi_1 T\,,
\end{equation}
where the respective Quadratic $\L_2$, Cubic $\L_3$, Quartic $\L_4$ and Quintic Galileon $\L_5$ interactions are given by
\ba
\L_n(\pi_1,\cdots,\pi_N) = \sum_{ m_1+\dots+m_N = n-1}\mathcal{L}_{m_1,\dots,m_N}
\ea
with
\begin{equation}
\label{MultiLag}
\mathcal{L}_{m_1,\dots,m_N}=(\alpha^1_{m_1,\dots,m_N}\pi_1+\cdots+\alpha^N_{m_1,\dots,m_N}\pi_N)\mathcal{E}_{m_1,\dots,m_N}\,,
\end{equation}
where the $\alpha^n_{m_1,\dots,m_N}$ are the coefficients for the Galileon interactions. Notice that this parameterization allow for  a lot of redundancy, so not all the $\alpha^n_{m_1,\dots,m_N}$ are meaningful (many of them can be set to zero without loss of generality). Notice as well that in this language these coefficients $\alpha$'s are dimensionfull (the dimension depends on $m_1+\dots+m_N$). We stick nonetheless to this notation for historical reasons, \cite{Padilla:2010de}.
In this formalism, all the derivative are included in the $\mathcal{E}_{m_1,\dots,m_N}$ which can be expressed as
\begin{eqnarray}
\mathcal{E}_{m_1,\dots,m_N}&=&(m_1+\cdots+m_N)!\delta^{\mu_1}_{[\nu_1}\cdots\delta^{\mu_{m_1}}_{\nu_{m_1}}
\cdots\delta^{\rho_1}_{\sigma_1}\cdots\delta^{\rho_{m_N}}_{\sigma_{m_N}]}\\ \nonumber
&& \times\left[(\partial_{\mu_1}\partial^{\nu_1}\pi_1)\cdots(\partial_{\mu_{m_1}}
\partial^{\nu_{m_1}}\pi_1)\right]\cdots\left[(\partial_{\rho_1}\partial^{\sigma_1}\pi_N)
\cdots(\partial_{\rho_{m_N}}\partial^{\sigma_{m_N}}\pi_N)\right]\,,
\end{eqnarray}
using the formalism derived in Ref.~\cite{Deffayet:2009mn}.

\paragraph{Bi-Galileon:}
Specializing this to the Bi-Galileon $N=2$ is straightforward. The analogue to \eqref{MultiLag} for the Bi-Galileon would simply be
\begin{equation}
\mathcal L_{\pi_1,\pi_2}=\sum_{0\leq m+n\leq 4}(\alpha_{m,n}\pi_1+\beta_{m,n}\pi_2)\mathcal E_{m,n}\,,\label{BiLag}
\end{equation}
with
\begin{equation}
\mathcal E_{m,n}=(m+n)!\delta^{\mu_1}_{[\nu_1}\cdots\delta^{\mu_m}_{\nu_m}\delta^{\rho_1}_{\sigma_1}
\cdots\delta^{\rho_m}_{\sigma_n]}(\partial_{\mu_1}\partial^{\nu_1}\pi_1)
\cdots(\partial_{\mu_m}\partial^{\nu_m}\pi_1)(\partial_{\rho_1}\partial^{\sigma_1}\pi_2)
\cdots(\partial_{\rho_n}\partial^{\sigma_n}\pi_2)\,.\label{EBi}
\end{equation}
The equations of motion for the two scalar fields $\pi_1$ and $\pi_2$ are then
\begin{equation}
\sum_{0\leq m+n\leq4}a_{m,n}\mathcal E_{m,n}=-T\;\;\;\;\mbox{and}\;\;\;\;\sum_{0\leq m+n\leq4}b_{m,n}\mathcal E_{m,n}=0\,,
\end{equation}
where the coefficients $a_{m,n}$ and $b_{m,n}$ can be expressed in terms of the parameters $\alpha_{m,n}$ and $\beta_{m,n}$ as
\begin{equation}
a_{m,n}=(m+1)(\alpha_{m,n}+\beta_{m+1,n-1})\;\;\;\;\mbox{and}\;\;\;\;b_{m,n}=(n+1)(\beta_{m,n}+\alpha_{m-1,n+1})\,.
\end{equation}

\section{Spherical symmetric backgrounds}\label{pert}
In this subsection, we recapitulate the formalism needed to study the superluminality of fluctuations  about spherical symmetric solutions. For this we split every field into a spherically symmetric background configuration $\pi^0(r)$ and a fluctuation $\delta \pi(t, \vec r)$,
\ba
\pi_n(t, \vec{r})=\pi^0_n(r)+\delta\pi_n(t, \vec{r})\,, \hspace{20pt}\forall\hspace{5pt}n=1,\cdots, N\,,
\ea
and introduce the $N$-dimensional fluctuation vector in field space,
\begin{equation}
\delta\Pi(t, \vec{r})=\begin{pmatrix}
\delta\pi_1(t, \vec{r})\\ \vdots  \\ \delta\pi_N(t, \vec{r})
\end{pmatrix}\,.
\end{equation}
At quadratic order in the fluctuations, the Lagrangian can be written as
\begin{equation}
\L_{\pi_1,\dots,\pi_N}=\frac{1}{2}\partial_t \Pi \, .\, \mathcal{K} \, .\,  \partial_t \Pi -\frac{1}{2} \partial_r \Pi  \, .\, \mathcal{U}  \, .\,  \partial_r\Pi -\frac{1}{2}\partial_\Omega \Pi \, .\,  \mathcal{V}  \, .\,  \partial_\Omega \Pi\,. \label{lag}
\end{equation}
The kinetic matrix $\mathcal{K}$  and the two gradient matrices
$\mathcal{U}$ and $\mathcal{V}$ are defined as follows:
\begin{align}
\label{eq:K-def}
\mathcal{K}&= (1+\frac{r}{3}\partial_r) \left(\Sigma_1+3 \Sigma_2 + 6\Sigma_3 + 6 \Sigma_4 \right)\\
\label{eq:U-def}
\mathcal{U}&=\Sigma_1+2\Sigma_2+2\Sigma_3\\
\label{eq:V-def}
\mathcal{V}&=(1+\frac{r}{2}\partial_r)\mathcal{U}\,,\
\end{align}
where the $\Sigma$ matrices depend on the spherically symmetric background configuration (and are thus functions of $r$).  In this language the $n^{\rm th}$ matrix $\Sigma_n$ encodes information about the $(n+1)^{\rm th}$ order Galileon interactions $\L_{n+1}$,
\begin{equation}
\Sigma_n=\begin{pmatrix}
\partial_{y_1}f_n^{a_1} &\cdots& \partial_{y_N} f_n^{a_N} \\
\vdots &\ddots& \vdots\\
\partial_{y_1} f_n^{a_1} &\cdots& \partial_{y_N} f_n^{a_N}
\end{pmatrix}\,,
\end{equation}
with
\ba
 f_n^\alpha(y_1(r),\cdots,y_N(r))=\sum_{i=0}^n (\alpha^{1'}_{i,n-i}+\alpha^{N'}_{i,n-i})y_1^i(r)\cdots y_N^{n-i}(r)\,,
\ea
and for each of the Galileon field, we define,
\ba
y_n(r)=\frac{\p_r \pi^0_n(r)}{r}\,.
\ea
In terms of the matrix $\mathcal{U}$, the background equations of motion are given by
\ba
\frac 1{r^2} \p_r\(r^2 \mathcal{U}(\pi^0(r)) . \p_r\(\begin{array}{c}\pi^0_1(r) \\ \pi^0_2(r) \\ \vdots \\ \pi^0_N(r) \end{array}\)\)=-
\(\begin{array}{c}T \\ 0 \\ \vdots \\ 0 \end{array}\)\,.
\ea
In particular for a point source of mass $M=4 \pi m$ localized at the origin $r=0$, we have
\ba
\label{eq:Multi_eom}
\(\Sigma_1+2\Sigma_2(r)+2\Sigma_3(r)\) . \(\begin{array}{c}y_1(r) \\ y_2(r) \\ \vdots \\ y_N(r) \end{array}\)=
\(\begin{array}{c}\frac{m}{r^3} \\ 0 \\ \vdots \\ 0 \end{array}\)\,,
\ea
where $\Sigma_1$ is independent of $y_n$ and is thus simply a constant, $\Sigma_2$ is linear in the $y_n$ and $\Sigma_3$ is quadratic in the fields.

Notice that the expressions (\ref{eq:K-def}, \ref{eq:U-def}, \ref{eq:V-def}) for the matrices $\mathcal{K}$, $\mathcal{U}$ and $\mathcal{V}$ in terms of the $\Sigma_n$ matrices are universal and do not depend on the number $N$ of fields.

\paragraph{Focus on the Bi-Galileon:}$\phantom{.}$\\
In the following we restrict our attention to the Bi-Galileon since we will first focus on that case and then generalize our results to the Multi-Galileon case. In the Bi-Galileon case, the matrices $\Sigma_n$ are given explicitly as below:
\begin{equation}
\Sigma_n=\begin{pmatrix}
\partial_yf_n^a & \partial_y f_n^b \\ \partial_z f_n^a & \partial_z f_n^b
\end{pmatrix} \;\;\;\;\;\;\;\;\;\;\mbox{with} \;\;\;\;\;\;\;\;\;\; f_n^\alpha=\sum_{i=0}^n \alpha'_{i,n-i}y^iz^{n-i}
\end{equation}
The functions $y$ and $z$ appearing in the coefficients $f_n^\alpha$ are shortcuts for
\ba
y(r)&=&\frac{1}{r}\frac{\partial \pi^0_1}{\partial r}\\
z(r)&=&\frac{1}{r}\frac{\partial \pi^0_2}{\partial r}\,,
\ea
 such that the equations of motion for $\pi_1^0$ and $\pi_2^0$ become simply
\begin{align}
f_1^a+2(f_2^a+f_3^a)&=\frac{m}{r^3}\\
f_1^b+2(f_2^b+f_3^b)&=0\,,
\end{align}
where $m=M/4\pi$, and $M$ is the mass of the point particle introduced at $r=0$.
More explicitly, in terms of $y$ and $z$ the two equations of motion are given by
\ba
a_{10}y+a_{01}z&+& 2\left(a_{20} y^2 + a_{11} y z + a_{02} z^2\right)\nonumber\\
&+& 2\left(+a_{30}y^3+a_{21}y^2z+a_{12}yz^2+a_{03}z^3\right)=\frac{m}{r^3}, \label{EOM1}\\
a_{01}y+b_{01}z&+&2\left(\frac{a_{11}}{2} y^2 + 2a_{02}y z + b_{02} z^2\right)\nonumber\\
&+&2\left(a_{21}/3y^3+a_{12}y^2z+3a_{03}yz^2+b_{03}z^3\right)=0\,. \label{EOM2}
\ea
In terms of the parameters $a_{ij}$ and $b_{ij}$,  the $\Sigma_{1,2,3}$ matrices can then be expressed respectively as
\ba
\label{Sigma1}
\Sigma_1&=&\begin{pmatrix}
a_{10}\  & \ a_{01}  \\ a_{01} \  & \  b_{01}
\end{pmatrix},\\
\label{Sigma2}
\Sigma_2&=&\begin{pmatrix}
2 a_{20} y + a_{11} z  \  & \  a_{11} y + 2 a_{02} z  \\ a_{11} y + 2 a_{02} z \  & \  2 a_{02} y + 2 b_{02} z
\end{pmatrix},\\
\label{Sigma3}
\Sigma_3&=&\begin{pmatrix}
3a_{30}y^2+2a_{21}yz+a_{12}z^2 \  & \  3a_{03}z^2+2a_{12}zy+a_{21}y^2  \\ 3a_{03}z^2+2a_{12}zy+a_{21}y^2 \  & \ 3b_{03}z^2+6a_{03}zy+a_{12}y^2
\end{pmatrix}\,,
\ea
in the Bi-Galileon case.
To get these expressions we have used the fact that for $m\textless n$ we have the correspondences $\mathcal E_{m,n}=\mathcal E_{n,m}|_{\pi_1^0\leftrightarrow\pi_2^0}$.
The exclusion of superluminal mode propagation implies that the sound speed of both modes along both the radial and orthoradial directions be less than or exactly equal to $1$. The two sound speeds in the radial direction are given by the eigenvalues of the matrix $\mathcal{M}_r=\mathcal{K}^{-1}\mathcal{U}$
and the two sound speeds along the orthoradial direction are given by the eigenvalues of the matrix $\mathcal{M}_\Omega=\mathcal{K}^{-1}\mathcal{V}$.
Therefore the condition for no superluminality is equivalent to requiring that all the eigenvalues of both matrices $\mathcal M_r-\mathbb{I}$ and $\mathcal M_\Omega-\mathbb{I}$ be zero or negative (and larger than $-1$), with $\mathbb{I}$ the identity matrix. In the following sections we study the behavior of the system in two different regimes, in the large and short distance regimes and check explicitly there always exists at least one superluminal mode in one direction.


\section{Superluminalities at Large Distances} \label{large-distance}

Let us start by summarizing the results find in this section:
 \begin{itemize}
 \item We first show that if at least one Cubic Galileon interaction is present\footnote{It can be an interaction involving just one of the $N$ Galileon fields, or an interaction mixing different Galileon fields together, the result remains unchanged.} \ie if $\L_3$ does not vanish identically, then superluminal propagation is always present at large enough distances from a point source. The only way to bypass this conclusion is to remove all the Cubic Galileon interactions for all $N$ fields $\L_3 \equiv 0$, meaning that any $\alpha^n_{m_1,\cdots,m_N}$ with $n=1,\cdots,N$ and $m_1+\cdots+m_N=2$ has to vanish exactly (for example in the Bi-Galileon case, this implies $a_{20}=a_{11}=a_{02}=b_{02}=0$). If the coefficients are merely small, then one can always go to large enough distances where the Cubic Galileon dominates over the Quartic and Quintic Galileon interactions.

\item Nevertheless even if all the Cubic Galileon terms vanish $\L_3 \equiv 0$, we can still find perfectly sensible static, spherically symmetric matter distributions around which there are superluminalities due to the Quartic Galileon at large distances.

\item As a consequence, we will see in this section that as soon as either a Cubic or a Quartic Galileon interaction is present one can always construct a sensible matter distribution which forces at least one of the $N$ Galileon fields to propagate superluminaly in one direction (either the radial or the orthoradial one).

\end{itemize}
We note that the Quintic Galileon interactions $\L_5$ always vanish at the background level around static, spherically symmetric sources, independently of the number of fields, so that if one tries to avoid the above conclusions by making both the Cubic and the Quartic Galileon vanish, then there is no Vainshtein mechanism at all about these configurations.

\subsection{Superluminalities from the Cubic Galileon}\label{CubicGalrlarge}

In the Multi-Galileon case, the background equations of motion for a point source at $r=0$ are given in \eqref{eq:Multi_eom}. At  large distances\footnote{We assume throughout this manuscript trivial asymptotic conditions at infinity which implies that the Galileon interactions ought to die out at large distances. The contributions from $\Sigma_1$ are thus the leading ones at large distances. Consistency of the theory requires that $\det \Sigma_1 \ne 0$ (so that the theory does indeed exhibit $N$ degrees of freedom) and the matrix $\Sigma_1$ is thus invertible.}, this simplify to
\ba
\label{eq:firstOrderInfinity}
\Sigma_1 \, . \(\begin{array}{c}y_1(r) \\ y_2(r) \\ \vdots \\ y_N(r) \end{array}\)=
\(\begin{array}{c}\frac{m}{r^3} \\ 0 \\ \vdots \\ 0 \end{array}\)\,.
\ea
Recalling that $\Sigma_1$ is invertible and independent of the field ($\Sigma_1$ is a constant), this implies that to leading order at large distance about a point-source, the fields die off as $r^{-1}$
\ba
y(r) \sim r^{-3}+ \mathcal{O}(r^{-6}) \hspace{20pt}\Rightarrow \hspace{20pt} \pi^0(r)\sim r^{-1}+\mathcal{O}(r^{-4})\,,
\ea
for at least one of the $N$ fields, as expected from the Newtonian inverse square law which should be valid at infinity.
As a result, at large distances the $\Sigma$ matrices behave as follows:
\ba
\Sigma_1&=& \bar \Sigma_1\\
\Sigma_2&=& \frac{1}{r^3}\bar \Sigma_2+ \mathcal{O}\(r^{-6}\)\\
\Sigma_3&=& \frac{1}{r^6}\bar \Sigma_3+ \mathcal{O}(r^{-9})\,,
\ea
where the `barred' matrices $\bar \Sigma_{1,2,3}$ are independent of $r$.

As a direct result of this scaling, it is trivial to see that at large distances, the kinetic and gradient matrices $\mathcal{K}$, $\mathcal{U}$ and $\mathcal{V}$ are given by
\begin{eqnarray}
\mathcal{K}&=&\bar\Sigma_1+0+\mathcal{O}\(\frac{1}{r^6}\) \,,\\
\mathcal{U}&=&\bar \Sigma_1+\frac{2}{r^3}\bar \Sigma_2+\mathcal{O}\(\frac{1}{r^6}\)\\
\mathcal{V}&=&\bar \Sigma_1-\frac{1}{r^3}\bar \Sigma_2 +\mathcal{O}\(\frac{1}{r^6}\)\,.
\end{eqnarray}
It is apparent that the perturbations at the order $\frac{1}{r^3}$ in the matrix $\mathcal{K}$ vanish while in the matrices $\mathcal{U}$ and $\mathcal{V}$ they always come with the opposite sign, hence there is always a superluminal direction at infinity. These results coincide with what is already known in the case of one Galileon \cite{Nicolis:2008in}. This is intrinsic to the $\frac{1}{r^3}$ behaviour at infinity and to the presence of the Cubic Galileon, and is independent of the number of fields.

The only way to bypass this very general result is to require the matrix $\bar \Sigma_2$ to vanish entirely, which
could be for instance achieved by imposing all the Cubic Galileon interactions to vanish\footnote{At large enough distances the Cubic Galileon would always dominate over the Quartic one (assuming trivial asymptotic conditions at infinity), so imposing a hierarchy between the Cubic Galileon interactions and the other ones is not sufficient to avoid superluminalities. All the Cubic interactions should be completely absent.}.
In particular, even if some eigenvalues of $\bar \Sigma_2$ vanish, the previous result remains unchanged, as long as $\bar \Sigma_2$ has at least one non-vanishing eigenvalue which would imply that the associated eigenmode in field space has a superluminal direction (either a radial or an orthoradial one). Only if all the eigenvalues of $\bar \Sigma_2$ vanish can we evade the previous argument, which can be accomplished by demanding all the coefficients arising from the cubic Galileon interactions to vanish exactly, in other words if the next to leading interactions arise from the Quartic Galileon.

In the very special case where  $\bar\Sigma_2$  vanishes entirely (\ie all its eigenvalues are identically $0$), then the previous argument needs special care.
The contribution from $\bar\Sigma_3$ implies that $(\mathcal{K}^{-1} \mathcal{U})$ and $(\mathcal{K}^{-1}\mathcal{V})$ do not necessarily have opposite sign. In any Multi-Galileon theory one can always tune the coefficients of  $\mathcal{L}_3$ so that the matrix $\bar \Sigma_2$ vanishes identically and so the $r^{-3}$ scaling is not the leading order correction to $\mathcal{U}$ and $\mathcal{V}$. For example, for the Bi-Galileon if the parameters of the theory are carefully chosen so as to satisfy 
$a_{20}=b_{02}  c^3$, $a_{11}=2 b_{02} c^2$ and $a_{02}=b_{02} c$ with $c=a_{01}/b_{01}$
then $\mathcal{U}$ and $\mathcal{V}$ vanish identically at $\mathcal{O}(r^{-3})$ for a pure point source and the argument given above breaks down. However as soon as we consider an extended source with energy density going as $1/r^{3-\epsilon}$ would revive $\bar\Sigma_2$ and the argument would then again be the one above. So even for these special coefficients, there is a whole classs of otherwise physically sensible solutions which exhibit superluminal propagation. 

In section \ref{QuarticGalileonsection} we consider this case more closely in the Bi-Galileon scenario and find that there is still always at least a superluminal mode for some range of $r$. For instance superluminalities unavoidably arise near the origin through the Quartic Galileon, unlike what was claimed in \cite{Padilla:2010tj}. But first we point out that one can very easily construct an extended source for which superluminalities are present at large distance for the Quartic Galileons just in the same way as they were for the Cubic ones.

\subsection{Superluminalities from the Quartic Galileon about an Extended Source}

When the Cubic Galileon is absent, the presence of superluminalities about a point-source is more subtle to prove and will be done explicitly in the next section.
Nevertheless, even if the coefficients in the Cubic Galileon vanish, we can always find a background configuration in which we can see superluminalities at large distances arise using the same argument as for the Cubic Galileon. In particular we can consider a gas of particles with a spatially varying density of the form
\begin{equation}\label{eq:l4-source}
T=M_0 \left(\frac{r_0}{r}\right)^{3/2}\,
\end{equation}
where $r_0$ characterizes the scale over which the density varies and $M_0$ controls the overall strength of the density profile\footnote{This matter  distribution can always be imagined for some arbitrarily large radius before being cut off.}.

In this case the asymptotic behavior of the background fields becomes
\begin{align}
\label{eq:y-large-r}
y_n(r)&=\frac{Y^{(1)}_n}{r^{3/2}}+\frac{Y^{(2)}_n}{r^{9/2}}+\mathcal{O}(\frac{1}{r^{15/2}})\,,
\end{align}
for all the fields $n=1, \cdots, N$, and we find once again using Eq.~\eqref{eq:K-def} that to order $\mathcal{O}(r^{-3/2})$, $\mathcal{K}$ vanishes and $\mathcal{U}$ and $\mathcal{V}$ have opposite signs, guaranteeing a superluminal direction.

This illustrates the basic reason we expect any theory that exhibits the Vainshtein mechanism to inevitably contain superluminalities when considering trivial asymptotic conditions. Every new source configuration gives rise to a new background Galileon field configuration. Because the theory must be nonlinear in order to have a Vainshtein mechanism, the fluctuations around this background will propagate on an effective metric determined by the background. Since the sources are not constrained by the theory, we are free to choose any source we like, and so we have a lot of freedom to change the parameters in this background metric and create superluminalities.

\section{Quartic Galileon about a point-source}\label{QuarticGalileonsection}

In the previous section we have shown that superluminalities are ubiquitous in Galileon models. No matter the number of field there is no possible choice of parameters that can ever free the theory from superluminal propagation. The argument in the previous section was completely generic an independent of the number of fields. It only required the behaviour at large distances (when the non-linear Galileon interactions can be treated perturbatively).

In what follows we show that even for a point source in the Quartic Bi-Galileon model\footnote{\ie we only consider two fields with mixing kinetic terms and Quartic Galileon interactions, but no Cubic interactions.}, superluminalities can never be avoided in a consistent model\footnote{The only requirements are the absence of ghost, the presence of an active Vainshtein mechanism, and trivial conditions at infinity.}. This result is in contradiction with previous results and examples in the literature, but upon presentation of this following argument, the previous claims have been reconsidered.

The philosophy of the argument goes as follows: We analyze the model both at large distances in the weak field limit and at short distances from the point source where the quartic interactions dominate (as required by the existence of an active Vainshtein mechanism). The requirement for stability (in  particular the absence of ghost) sets some conditions on the parameters of the theory. We then show that these conditions are sufficient to imply the presence of superluminal modes near the source.
We emphasize that this result could not be obtained, should we just have focused on the near origin behaviour without knowledge of the field stability at infinity.

\subsection{Stability at Large Distances}

To ensure the stability of the fields, the kinetic matrix $\mathcal{K}$ as well as the gradient matrices $\mathcal{U}$ and $\mathcal{V}$ should be positive definite at any point $r$. At infinity in particular these three conditions are equivalent and simply imply that the matrix $\Sigma_1$ ought to be positive definite. In the case of the Bi-Galileon, this implies
\begin{eqnarray}
\det \Sigma_1=a_{10}b_{01}-a_{01}^2>0\label{cond}\;\;\;\;\;\mbox{and}\;\;\;\;\;{\rm Tr}\ \Sigma_1=a_{10}+b_{01}>0\,.
\end{eqnarray}
In terms of the coefficients of the quadratic terms, these two conditions imply
\begin{eqnarray}
a_{10}>0\;\;\;\;\;\mbox{and}\;\;\;\;\;b_{01}>\frac{a_{01}^2}{a_{10}}\, >\, 0\,. \label{b10}
\end{eqnarray}
The behaviour of the fields at large distance from a point source localized at $r=0$ is determined by the coefficients of the quadratic terms, (or equivalently by $\Sigma_1$),
\begin{align}
\label{eq:Y1Z1}
y(r)&=\frac{Y_1}{r^3}+\mathcal{O}\(\frac{1}{r^6}\)\,,\;\;\;\;\;\;\mbox{with}\;\;\;\;\;\;Y_1=\frac{b_{01}}{\mbox{det}\Sigma_1}m\\
z(r)&=\frac{Z_1}{r^3}+\mathcal{O}\(\frac{1}{r^6}\)\,, \;\;\;\;\;\;\mbox{with}\;\;\;\;\;\;Z_1=\frac{-a_{01}}{\mbox{det}\Sigma_1}m\,,
\end{align}
and as expected, we recover a Newtonian inverse square law behavior for each mode at infinity,
namely $\partial_r \pi_1^0 \sim \partial_r \pi_2^0 \sim r^{-2}$. At this stage it is worth to mention that the stability condition \eqref{b10} implies that $Y_1>0$, which is consistent with the fact that the  force mediated by the one field $\pi_1$ that coules to matter is attractive.

\subsection{Short Distance Behavior}

We now study the field fluctuations at small distances near the source (\ie at leading order in $r$, assuming we are well within the Vainshtein region).
 From the equations of motion \eqref{EOM1}, \eqref{EOM2} after setting the coefficients of the cubic Galileon to zero) near the  origin, we infer the following expansion
\begin{align}
y(r)&=\frac{y_1}r+y_2 r+\O(r^3)\\
z(r)&=\frac{z_1}r+z_2r+\O(r^3)\,,
\end{align}
with
\begin{eqnarray}
a_{30}y_1^3+a_{21}y_1^2z_1+a_{12}y_1z_1^2+a_{03}z_1^3&=&\frac{m}{2}\label{ordre1_1}\\
\frac{a_{21}}{3}y_1^3+a_{12}y_1^2z_1+3a_{03}y_1z_1^2+b_{03}z_1^3&=&0\label{ordre1_2}\,.
\end{eqnarray}
Note that the $O(r^0)$ terms vanish since the Cubic Galileon is not present.\\
Expanding $\Sigma_3$ in powers of $r$, we have
\begin{equation}
\Sigma_3=\Sigma_3^{(l)}+\Sigma_3^{(nl)}+\cdots=\frac{1}{r^2}\tilde{\Sigma}_3^{(l)}+\tilde{\Sigma}_3^{(nl)}+\mathcal{O}\(r^2\)\,,
\end{equation}
where the leading order contribution to $\Sigma_3$ is given by
\begin{equation}
\tilde{\Sigma}_3^{(l)}=\begin{pmatrix} 3a_{30}y_1^2+2a_{21}y_1z_1+a_{12}z_1^2\ & \ 3a_{03}z_1^2+2a_{12}y_1z_1+a_{21}y_1^2 \\3a_{03}z_1^2+2a_{12}y_1z_1+a_{21}y_1^2\  &\  3b_{03}z_1^2+6a_{03}y_1z_1+a_{12}y_1^2 \end{pmatrix}\,.
\end{equation}
Solving the equation of motion (\ref{ordre1_2}) for $b_{03}$ gives $b_{03}=(-a_{21}y^3-3a_{12}y^2z-9a_{03}yz^2)/(3 z^3)$. Similarly solving the equation of motion (\ref{ordre1_1}) for $a_{30}$ yields $a_{30}=((m - 8 a_{21} \pi r^3 y^2 z - 8 a_{12} \pi r^3 y z^2 - 8 a_{03} \pi r^3 z^3)/(8 \pi r^3 y^3))$. Using these expressions for $b_{03}, a_{30}$ and introducing the combination $\mathcal{B}$ defined as follows:
\ba
\mathcal{B}=3a_{03}z_1^2+2a_{12}y_1z_1+a_{21}y_1^2\,,
\ea
we can then write $\tilde{\Sigma}_3^{(l)}$ simply as :
\begin{equation}
\label{Sigma3_l}
\tilde{\Sigma}_3^{(l)}=\begin{pmatrix} -\frac{z_1}{y_1}\mathcal{B}+\frac{3m}{2y_1} & \mathcal{B} \\ \mathcal{B}  &-\frac{y_1\mathcal{B}}{z_1} \end{pmatrix}\,.
\end{equation}

\subsection{Stability at Short Distances}

As mentioned previously, me must ensure that the eigenvalues of $\mathcal{K}$ are strictly positive.  At small distances near the source,  the matrix $\mathcal{K}$ can be expressed as
\begin{equation}
\mathcal{K}=\frac{2}{r^2}\tilde{\Sigma}_3^{(l)}+\O(1)\,.
\end{equation}
In terms of $y_1$, $z_1$ and $\mathcal{B}$, the absence of ghost near the origin implies the following conditions on the parameters
\begin{align}
\mbox{det}\tilde\Sigma_3=-\frac{3}{2}\frac{\mathcal{B}m}{z_1}&>0 \\
\mbox{Tr}\tilde\Sigma_3=\frac{3 m z_1 - 2 \mathcal{B} (y_1^2 + z_1^2)}{y_1 z_1}&>0\,,
\end{align}
which are equivalent to
\ba
\label{y1}
y_1>0,\;\;\;\;\;\mbox{and}\;\;\;\;\;\;\frac{\mathcal{B}}{z_1}<0\,.
\label{B}
\ea
We now use the stability conditions derives at both large and small distances to deduce the behaviour of the radial sound speed in this model.

\subsection{Sound Speed near the Source}
Similarly as we did at large distances, we can now compute the `radial sound speed' matrix $\mathcal{M}_r=\mathcal{K}^{-1}\mathcal{U}$ near the origin,
\begin{equation}\label{eq:M-leading-order}
\mathcal{M}_r =\mathbb{I}-2r^2(\tilde{\Sigma}_3^{(l)})^{-1}\tilde{\Sigma}_3^{(nl)}+\O(r^4)\,.
\end{equation}
We note that unlike the Cubic Galileon case described in more detail below, the leading order behaviour of $\mathcal{M}$ is not manifestly superluminal. However, this is not enough to guarantee the absence of superluminal modes, we must carefully check the sign of the small $\mathcal{O}(r^2)$ correction term before making any conclusions.
A simple formulation for the matrix $\Sigma_3^{(l)}$ is given in \eqref{Sigma3_l}, and a similar expression for $\Sigma_3^{(nl)}$ can be found in an analogous way,
\begin{equation}
\tilde{\Sigma}_3^{(nl)}=\begin{pmatrix} -a_{10}-a_{01}\frac{z_1}{y_1}-2\zeta\frac{z_1}{y_1} & 2\zeta \\ 2\zeta & -b_{01}-a_{01}\frac{y_1}{z_1}-2\zeta\frac{y_1}{z_1} \end{pmatrix}\,,
\end{equation}
with the notation:
\begin{equation}
\zeta=a_{21} y_1 y_2 + a_{12} y_2 z_1 + a_{12} y_1 z_2 + 3 a_{03} z_1 z_2\,.
\end{equation}
This allows us to compute the radial sound speed\footnote{We need to work to $\O(r^4)$ inside the square root to get this expression, because we need to square the $\O(r^2)$ corrections we have calculated. One might worry that the calculation we have done is not consistent because we have not worked to $\O(r^4)$, however one can check that the $\O(r^4)$ corrections we have neglected here cancel identically and do not contribute to $c_s^2$ at the order we are interested in.}:
\begin{equation}
{c_s^2}_\pm =1+r^2(a'\pm\sqrt{b'})+\O(r^4)\,,
\end{equation}
with $a'$ and $b'$ some coefficients that depend on $y_1$, $z_1$, $\mathcal{B}$, $m$ and $(a,b)_{ij}$. So for both modes to be subluminal along the radial direction, the following conditions should be satisfied:
\ba
a'<0,
\hspace{25pt}b'>0\hspace{15pt}{\rm and}\hspace{15pt}
a'^2-b'>0\,.
\ea
However as we shall see, these are not consistent with the stability conditions established previously.

The explicit form of the coefficients $a'$ and $b'$\footnote{It easy to check that $b'$ is always positive.} is given by :
\ba
a'&=&- \frac{1}{3m \mathcal{B}y_1}\Big(3 m(a_{01} y_1 + 2 \zeta y_1 + b_{01} z_1)-2 \mathcal{B} \mathcal{D}  \Big)\\
a'^2-b'&=&-\frac{8}{3 m \mathcal{B}y_1}\Big( (a_{10}y_1+a_{01}z_1)(a_{01}y_1+b_{01}z_1)+2\zeta \mathcal{D}  \Big)\label{condImp}\,,
\ea
with the notation
\ba
\mathcal{D}=a_{10}y_1^2+2a_{01}z_1y_1+b_{01}z_1^2\,.
\ea
We may re-express the quantity $\mathcal{D}$ as follows
\ba
\mathcal{D}&=&a_{10} y_1^2 + 2 a_{01} z_1y_1 + b_{01} z_1^2\\
&=&a_{10}\(y_1+\frac{a_{01}}{a_{10}}z_1\)^2+\frac{z_1^2}{a_{10}}\(a_{10}b_{01}-a_{01}^2\)>0\,.
\ea
Recall from the expression of the kinetic matrix $\K$ at infinity, that the following two conditions should be satisfied,
(\ref{cond}):
$a_{10}>0$ and $\(a_{10}b_{01}-a_{01}^2\)>0$, which directly implies that $\mathcal{D}$ is strictly positive.
Knowing this, we check whether or not $a'<0$ and $a'{}^2-b'>0$ which if true, would imply that both modes are sub-luminal.

We start with the requirement that  $a'<0$. This implies that $3m(a_{01} y_1 + 2 \zeta y_1 + b_{01} z_1)-2 \mathcal{B} (a_{10} y_1^2 + 2a_{01}z_1 y_1 + b_{01} z_1^2)$ has the same sign as $\mathcal{B}$. Once this condition is satisfied, we check the sign of $a'^2-b'>0$. This quantity is positive only if $\mathcal{F}$ has the opposite sign as $\mathcal{B}$, where
\begin{equation}
\mathcal{F}=(a_{10}y_1+a_{01}z_1)(a_{01}y_1+b_{01}z_1)+2\zeta \mathcal{D}\,.
\end{equation}
In what follows, we will start by assuming that $z_1$ is positive and show that in that case the condition to avoid any super-luminal modes cannot be satisfied. The same remains true if $z_1$ is assumed to be negative. We can therefore conclude that the Quartic Bi-Galileon interactions always produce a superluminal mode already in the configuration about a point source.

We recall from eq.~\eqref{B} that if $z_1>0$, the absence of ghost-like modes near the origin imposes the condition $\mathcal{B}<0$.
Furthermore, knowing that $\mathcal{D}=a_{10} y_1^2 + 2a_{01}z_1 y_1 + b_{01} z_1^2>0$, we can infer that $a'$ negative only if
\begin{equation}
 a_{01} y_1 + 2 \zeta y_1 + b_{01} z_1<\frac{2\mathcal{B}\mathcal{D}}{3m}<0\,.
\end{equation}
Then using the fact that $\mathcal{D}=a_{10} y_1^2 + 2a_{01}z_1 y_1 + b_{01} z_1^2>0$, this implies (knowing from \eqref{y1} that $y_1>0$):
\begin{eqnarray}
a_{01} y_1 + 2 \zeta y_1 + b_{01} z_1<0 \hspace{10pt}
\Rightarrow \hspace{10pt}\zeta<-\frac{1}{2}(a_{01}+b_{01}\frac{z_1}{y_1})\label{cond1}\,.
\end{eqnarray}
Finally to avoid any superluminal mode, the quantity $a'^2-b'$ should also be positive. Since in this case $\mathcal{B}$ is negative, $a'^2-b'$ has the same sign as $\mathcal{F}$, where
\ba
\mathcal{F}&=&(a_{10}y_1+a_{01}z_1)(a_{01}y_1+b_{01}z_1)+2\zeta \mathcal{D}\nn\\
&<&(a_{10}y_1+a_{01}z_1)(a_{01}y_1+b_{01}z_1)-(a_{01}+b_{01}\frac{z_1}{y_1})\mathcal{D}\nn\\
&<&-\frac{z_1}{y_1}\(a_{01}y_1+b_{01}z_1\)^2\,.
\ea
Since $y_1>0$ and $z_1>0$ this implies that $\mathcal F<0$. Since $a'^2-b'$ has the same sign as $\mathcal{F}$, we can therefore conclude that if we assume $z_1$ to be positive and $a'<0$,  the quantity $(a'^2-b')$ will also be negative, or in other words, there is one superluminal mode.
This argument was made assuming $z_1>0$, however it is straightforward to reproduce the same argument for negative $z_1$. If we choose for instance negative $z_1$ ($z_1<0$) then the condition coming from the absence of ghost-like instabilities eq.~\eqref{B} will require this time the opposite sign for $\mathcal{B}$, namely $\mathcal{B}>0$ and therefore $\mathcal{F}$ will be a positive number $\mathcal{F}<-\frac{z_1}{y_1}\(a_{01}y_1+b_{01}z_1\)^2$ while the expression $(a'^2-b')$ in eq.~\eqref{condImp} will have the opposite sign to $\mathcal{F}$ and therefore again there would not be any choice of coefficients $(a,b)_{ij}$ to make both modes (sub)luminal. With this we have proven that there is no generic choice for the parameters $a_{ij}$ and $b_{ij}$ near the origin which would prevent the propagation of superluminal modes.


\subsection{Special Case of Dominant First Order Corrections}

In the previous section we proved that close to the source there is always one mode which propagates superluminaly in a generic theory where only the Quartic Galileon is present. However we made a technical assumption in Eq.~(\ref{eq:M-leading-order}) that $\tilde{\Sigma}_3^{(l)}$ was invertible, or equivalently that we did not make the special choice of parameters $a_{ij},b_{0i}$ which gives $\mathcal{B}=0$ (implying that the leading order pieces in $\Sigma_3$ were strictly larger than the first order corrections). However we could consider a specific choice for which some of the leading order pieces of $\Sigma_3$ vanish and the subleading pieces become dominant. Therefore in this section we will examine this possible loophole more closely. We will find that even in this case one always finds that a superluminal mode is present. When $\mathcal{B}=0$, $\tilde{\Sigma}_3^{(l)}$ takes the following trivial form:
\begin{equation}
\tilde{\Sigma}_{3}^{(l)}=\left(\begin{array}{cc}
\frac{3m}{2y_{1}} & 0\\
0 & 0
\end{array}\right).
\end{equation}
The stability of the theory now depends not only on the leading behavior of the kinetic $\mathcal{K}$ and radial derivative $\mathcal{U}$ matrices, but also on the subleading behavior. In terms of the $\Sigma$ matrices, $\mathcal{K}$ and $\mathcal{U}$ take the following form
\begin{equation}
\mathcal{K}=2\frac{\tilde{\Sigma}_{3}^{(l)}}{r^{2}}+\Sigma_{1}+6\tilde{\Sigma}_{3}^{(nl)},\ \ \ \ \ \ \mathcal{U}=2\frac{\tilde{\Sigma_{3}}^{(l)}}{r^{2}}+\Sigma_{1}+2\tilde{\Sigma}_{3}^{(nl)}.
\end{equation}
The theory is stable only if  $\mathcal{K}$ and $\mathcal{U}$ have positive eigenvalues, or in other words only if the following quantities three quantities are positive:
\begin{eqnarray}
y_1&>&0,\nonumber\\
\lambda_{1}^{2}  \equiv -\frac{3m}{y_{1}z_{1}^{2}}\left(6(a_{01}+2\zeta)y_{1}z_{1}+5b_{01}z_{1}^{2}\right)&>&0,\\
\lambda_{2}^{2}  \equiv -\frac{3m}{y_{1}z_{1}^{2}}\left(2(a_{01}+2\zeta)y_{1}z_{1}+b_{01}z_{1}^{2}\right)&>&0.\nonumber
\end{eqnarray}
Now we construct again the radial speed of sound matrix $\mathcal{M}_r \equiv \mathcal{K}^{-1}\mathcal{U}$ in this specific case with $\mathcal{B}=0$.
We can write the trace and determinant as
\begin{eqnarray}
{\rm tr}\mathcal{M}_r  & = & \left(1+\frac{\lambda_{2}^{2}}{\lambda_{1}^{2}}\right)+r^{2}\tau+\O(r^{4}),\nonumber\\
{\rm det}\mathcal{M}_r  & = & \frac{\lambda_{2}^{2}}{\lambda_{1}^{2}}+r^{2}\delta+\O(r^{4}).
\end{eqnarray}
where $\tau$ and $\delta$ are functions of the given parameters (however we will only need $\tau-\delta$ as shown below).
The speed of sound is given by, to $\O(r^{2})$,
\begin{itemize}
\item If $\lambda_{1}^2>\lambda_{2}^2$
\end{itemize}
\begin{equation}
c_{\pm}^{2}=\begin{cases}
1+r^{2}\frac{\lambda_{1}^{2}}{\lambda_{1}^{2}-\lambda_{2}^{2}}(\tau-\delta),\\
\frac{\lambda_{2}^{2}}{\lambda_{1}^{2}}-r^{2}\frac{\lambda_{1}^{2}}{\lambda_{1}^{2}-\lambda_{2}^{2}}(\tau-\delta).
\end{cases}
\end{equation}
In this case we will have superluminal
propagation if and only if $\tau-\delta>0$. We show that one always has $\tau-\delta>0$ in this case by carefully making use of the stability constraints in Appendix~\ref{sec:app-b-equals-0}.

\begin{itemize}
\item If $\lambda_{1}^{2}<\lambda_{2}^{2}$
\end{itemize}
\begin{equation}
c_{\pm}^{2}=\begin{cases}
\frac{\lambda_{2}^{2}}{\lambda_{1}^{2}}+r^{2}\frac{\lambda_{1}^{2}}{\lambda_{2}^{2}-\lambda_{1}^{2}}(\tau-\delta),\\
1-r^{2}\frac{\lambda_{1}^{2}}{\lambda_{2}^{2}-\lambda_{1}^{2}}(\tau-\delta).
\end{cases}
\end{equation}
Since $\lambda_{1}^{2}<\lambda_{2}^{2}$, in this case
superluminal propagation is guaranteed.

One might argue that this only guarantees superluminalitiy at the origin which is inside the redressed strong coupling radius of the theory, where we can no longer trust the results of the theory. However, we note that explicitly factoring out powers of $M$ and $\Lambda$ that the speed of sound is given by
\be
c_{\pm}^{2}=\frac{\lambda_{2}^{2}}{\lambda_{1}^{2}}+\left(\frac{r}{r_{V}}\right)^{2}\frac{\lambda_{1}^{2}}{\lambda_{2}^{2}-\lambda_{1}^{2}}(\hat{\tau}-\hat{\delta}),
\ee
where $r_{V}\equiv(M/\mpl)^{1/3}/\Lambda$ is the Vainshtein radius and where $\hat{\tau}$ and $\hat{\delta}$ are dimensionless. Since $r_V>\Lambda^{-1}$, and since the redressed strong coupling radius is always smaller than $\Lambda^{-1}$, there is a range of $r$ in which we can trust the theory and we can also trust the leading order behavior of the speed of sound above.

\section{Cubic Lagrangian near the Source}

Lets have a quick look into the contributions coming from a Cubic Bi-Galileon theory\footnote{\ie a Bi-Galileon theory where only the Cubic interactions are present} near the origin and study the superluminality. In the section \ref{CubicGalrlarge} we had seen that the existence of the Cubic Galileon guarantees superluminal propagation at infinity. Now lets also see the effect of a pure Cubic Galileon term on short distances.
We quickly recall the equations of motion in the Cubic Galileon case near the origin here again:
\ba
2\left(a_{02}z^2+a_{11}yz+a_{20}y^2\right)&=&\frac{m}{r^3}\label{ecub0}\\
2\left(b_{02}z^2+2a_{02}yz+\frac12a_{11}y^2\right)&=&0\,.\label{ecub}
\ea
At short distance the fields then behave as
\ba
y(r)&=&\frac{y_1}{r^{3/2}}+y_2+\O(r^{3/2})\,,\\
z(r)&=&\frac{z_1}{r^{3/2}}+z_2+\O(r^{3/2})\,.
\ea
The leading order matrix $\tilde{\Sigma}_2^{(l)}$ can be expressed as follows (after use of the equations of motion):
\begin{equation}
\label{Sigma2_leom}
\tilde{\Sigma}_2^{(l)}=\begin{pmatrix} \frac{m}{y_1}-\frac{z_1}{y_1}\mathcal C & \mathcal C \\ \mathcal C  &-\frac{y_1}{z_1}\mathcal C \end{pmatrix}\,.
\end{equation}
with the notation
\begin{equation}
\mathcal C=a_{11}y_1+2a_{02}z_1\,.
\end{equation}
The stability condition in the short distance regime requires $\mbox{det}\mathcal K\approx\mbox{det}\Sigma^{(l)}_2>0$ and $\mbox{Tr}\mathcal K\approx\mbox{Tr}\tilde{\Sigma}^{(l)}_2>0$:
\begin{eqnarray}
\mbox{Tr}\Sigma_2^{(l)}=\frac{m-\mathcal C z_1}{y_1}>0\\
\mbox{det}\Sigma_2^{(l)}=\frac{-m\mathcal C}{z_1}>0\,.
\end{eqnarray}
These conditions imply $\frac{\mathcal C}{z_1}<0$ and $y_1>0$.
After using the next-to-leading order equations of motion to simplify the next-to-leading order matrix $\tilde{\Sigma}_2^{(nl)}$, we find
\begin{eqnarray}
\label{Sigma2_l}
\tilde{\Sigma}_2^{(nl)}=\begin{pmatrix} -\frac{a_{10}}{2}-a_{01}\frac{z_1}{2y_1}-2\beta\frac{z_1}{y_1} & 2\beta \\ 2\beta  &-\frac{b_{01}}{2}-\frac{a_{01}y_1}{2z_1}-2\beta\frac{y_1}{z_1} \end{pmatrix}\,,
\end{eqnarray}
with $\beta\equiv a_{02}z_2+\frac{a_{11}}{2}y_2$.

Assuming $\Sigma_{2}^{(l)}$ is invertible (which is the case if there is at least one non-vanishing Cubic Galieon interaction), the matrix $\mathcal M_r$ is given by
\begin{eqnarray}
\mathcal{M}_r & = & \mathcal{K}^{-1}\mathcal{U}=\left[\frac{3}{2}\frac{\Sigma_2^{(l)}}{r^{3/2}}+(\Sigma_{1}+3\Sigma_{2}^{(nl)})\right]^{-1}\left[2\frac{\Sigma_{2}^{(l)}}{r^{3/2}}+(\Sigma_{1}+2\Sigma_{2}^{(nl)})\right]\\
 & = & \frac{4}{3}-\frac{2}{3}r^{3/2}\left[\Sigma_{2}^{(l)}\right]^{-1}\left[\Sigma_{1}+4\Sigma_{2}^{(nl)}\right]\,.
\end{eqnarray}
This in turn implies that the Cubic Galileon also gives rise to superluminal propagation near the origin. If on  the other hand we consider  the possible loophole with vanishing determinant of the leading matrix $\Sigma_2^{(l)}$ (choosing parameters such $\mathcal{C}=0$) nothing changes. The matrix $\mathcal{M}_r $ has still one eigenvalue going as $4/3+\O(r)$, and another eigenvalue whose leading behavior depends on the signs and relative strengths of $\beta$ and $z_{1}$. But the existence of one eigenvalue that is $4/3$ at leading order is enough to prove the existence of superluminalities in that regime as well.


\section{Discussion}
In this paper we have shown that Multi-Galileon theories inevitably contain superluminal modes around some backgrounds, for any number of Galileon fields. At large distances from a static point source, we have shown that if the Cubic Galileon is present (even if its coefficients are very small), it will eventually dominate over the other Galileons and lead to a superluminal mode. Even if no Cubic Galileon interactions are  present (\ie all the Cubic Galileon coefficients are exactly zero), we find that there are simple, perfectly valid matter distributions (such as a static gas of particles whose density falls of as $r^{-3/2}$) around which perturbations propagate superluminally.

We also considered the case studied in the journal version of Ref.~\cite{Padilla:2010tj} of perturbations around a static point source in the Bi-Galileon when only the Quartic Galileon is present. By studying the speed of sound of perturbations close to the source, we find, in contradiction to their original claims, that the presence of a superluminal mode cannot be avoided. This is a nontrivial result, which can only be seen by carefully taking into account the constraints that stability at large distances places on the theory, and the interplay between these conditions coming from infinity and the action for perturbations near the source. In other words, this is not a local result which could have been derived from the knowledge of the behaviour near the source only. We have also showed that there will always be superluminal perturbations around a point source if only a Cubic Galileon is present.

Our results physically arise from the link between the Vainshtein mechanism and superluminalities in typical Galileon theories. So long as one is considering theories that are ghost-free, with trivial asymptotic conditions at infinity and avoid quantum strong-coupling issues (fields with vanishing kinetic terms), these two effects are intimately connected. One way to see this link is to note that the Vainshtein mechanism is inherently nonlinear, and so the behavior of the perturbations depends strongly on the source distribution present. Thus one expects to always be able to find backgrounds around which there are superluminalities. However, the connection may be stronger than this: As we have shown, even in the case of a static point source with only a Quartic Galileon present, where the presence of superluminalities at large distances is not manifest, there are still inevitably superluminalities close to the source.

We would like to emphasize however that the presence of superluminal modes is not enough to conclude that Galileon theories are inconsistent. As discussed in greater detail in Ref.~\cite{Burrage:2011cr}, the Galileons still have their own causal structure. The crucial issue is instead whether or not closed time like curves can form. This would lead to violations of causality and the theory would be inconsistent. However, a Chronology Protection Conjecture for Galileon theories can be constructed, which states that it is impossible to form closed timelike curves without requiring energies that push the theory beyond its regime of validity. Further work could explore this conjecture in greater detail.

We believe that the superluminalities are a crucial feature of Galileon theories. As shown in Ref.~\cite{Adams:2006sv}, the presence of superluminalities around some backgrounds is ultimately tied with the failure of the Galileon theory to have a Wilsonian completion. It would be interesting  to understand whether this aspect and the presence of a Vainshtein mechanism could however be tied to theories which allow for an alternative to UV completion such as classicalization, \cite{Dvali:2010jz,Dvali:2012zc,Vikman:2012bx}.

We conclude by reviewing the only known way (so far) to have a Vainshtein mechanism and still avoid superluminalities.  If the Galileon is not considered as a field in its own right, but rather as a component of anther fully fledge theory, one needs not to impose trivial asymptotic conditions at infinity. In massive gravity for instance, the Galileon field that appears in its decoupling limit is not a fundamental field. In such setups, it is then consistent to consider configurations for which the Galileon field does not vanish at infinity, so long as the metric is well defined at infinity (which does not necessarily imply Minkowski space-time).
In such cases, we can thus have more freedom to fix the asymptotic boundary conditions for the Galileon field. A specific realization has recently been found\footnote{In a single Galileon realization. Notice that the presence of multiple Galileon fields would not be relevant there.} in \cite{Berezhiani:2013dw}, where the asymptotic behaviour is non trivial and the metric asymptotes to a cosmological one at large distances. These results do rely on the existence of a non-trivial coupling to matter of the form $\p_\mu \pi \p_\nu \pi T^{\mu\nu}$ which naturally arises in Massive Gravity, \cite{deRham:2010ik}.
When  such non-trivial asymptotics conditions are considered, the results derived in this work are no longer valid and open the door for a way to find configurations which do exhibit the Vainshtein mechanism without necessarily propagating a superluminal mode around these configurations. Future work should consider the role of boundary conditions in the selection of viable configurations.

\section*{Comments}
Part of this work was been derived as Paul de Fromont's Master thesis in the summer of 2011 and was presented in Ref.~\cite{PaulThesis}. 

As a result to this work, the example and conclusions originally presented in Ref.~\cite{Padilla:2010tj} have been corrected, and a new arXiv version of \cite{Padilla:2010tj} has been submitted, taking into account and summarizing the new analysis performed here.

The results presented in this work are in agreement with that of Ref.~\cite{Garcia-Saenz:2013gya}. Whilst the methods used in \cite{Garcia-Saenz:2013gya} are different,  they reach the same basic conclusion that the combination of an active Vainshtein mechanism, a lack of ghost at infinity, and trivial boundary conditions at infinity inevitably lead to superluminalities in Multi-Galileon theories.

\acknowledgments{We would like to thank Sebastien Garcia-Saenz for useful discussions. 
PdF would like to thank the University of Geneva for its hospitality during the earlier part of this work.
AM would like to thank CERN for its hospitality during the latest stage of this work. AM is supported by the NSF GRFP program.}

\newpage

\appendix
%
%
%
%

\section{Detailed analysis of the Special Case in the Quartic Galileon: Dominant First Order Corrections}
\label{sec:app-b-equals-0}

In this section of the appendix we will show that $\tau-\delta>0$ when $\lambda_1^2>\lambda_2^2$.

We can write $\tau-\delta$ as
\begin{eqnarray}
\tau-\delta & = & \frac{6b_{01}m}{\lambda_{1}^{4}y_{1}}(1-\alpha)\nonumber
  \times  \Big[\frac{1}{8(3\alpha-1)^{2}}\left(a_{10}b_{01}-2\frac{3\alpha-1}{5\alpha-1}a_{01}^{2}\right)\\
  &+&\frac{2}{(1-\alpha)(5\alpha-1)}\zeta^{2}-\frac{1}{(5\alpha-1)(3\alpha-1)}\zeta a_{01}\Big]\,.
\end{eqnarray}
We also use the notation $\lambda_{2}^{2}=\alpha\lambda_{1}^{2}$.
Note that since $y_{1},b_{01}>0$ we must have $3\lambda_{2}^{2}>\lambda_{1}^{2}$. Thus we have $1/3<\alpha<1$, the upper bound comes from our assumption that $\lambda_1^2 > \lambda_2^2$.

Now the first term in the brackets of the expression $\tau-\delta$ has the same sign as
\begin{equation}
a_{01}b_{01}-\epsilon a_{01}^{2}\,,
\end{equation}
with $0<\epsilon<1$. This is positive because in order to avoid ghost instabilities at large distances from the source $a_{10}b_{01}-a_{01}^{2}>0$ and $a_{10},b_{01}>0$, (see equation (\ref{cond})). Meanwhile, the second term in the brackets is manifestly positive. Finally, the third term has the sign of $-\zeta a_{01}$.

So at this point our only hope of avoiding superluminalities is to consider a choice of parameters where $-\zeta a_{01}<0$. Now we will proceed to show  that $\tau-\delta>0$ in this case as well.

Note that in the limit $\alpha\rightarrow1$ with everything else
fixed we have
\begin{equation}
\tau-\delta\longrightarrow\frac{3b_{01}m\zeta^{2}}{\lambda_{1}^{4}y_{1}}>0\,.
\end{equation}
Also in the limit $\alpha\rightarrow1/3$ with everything else fixed
we have
\begin{equation}
\tau-\delta\longrightarrow\frac{a_{10}b_{01}^{2}m}{2\lambda_{1}^{4}y_{1}}\frac{1}{(1-3\alpha)^{2}}>0\,.
\end{equation}
Now consider the function
\begin{eqnarray}
\sigma(\alpha) & = & 8(1-\alpha)(5\alpha-1)(3\alpha-1)^{2}\Big[\frac{1}{8(3\alpha-1)^{2}}\left(a_{10}b_{01}-2\frac{3\alpha-1}{5\alpha-1}a_{01}^{2}\right)\nonumber\\
&+&\frac{2}{(1-\alpha)(5\alpha-1)}\zeta^{2}-\frac{1}{(5\alpha-1)(3\alpha-1)}\zeta a_{01}\Big]\,.
\end{eqnarray}
We can write this function in the shortened notation as $\sigma(\alpha)=\sigma_{0}+\sigma_{1}\alpha+\sigma_{2}\alpha^{2}$ with
\begin{eqnarray}
\sigma_{0} & = & 2a_{01}^{2}-a_{10}b_{01}+8a_{01}\zeta+16\zeta^{2}\\
\sigma_{1} & = & -8a_{01}^{2}+6a_{10}b_{01}-32a_{01}\zeta-96\zeta^{2}\\
\sigma_{2} & = & 6a_{01}^{2}-5a_{10}b_{01}+24a_{01}\zeta+144\zeta^{2}\,.
\end{eqnarray}
The sign of $\sigma(\alpha)$ is the same as the sign of $\tau-\delta$
in the regime $1/3<\alpha<1$. Note that $\sigma(1/3),\sigma(1)>0$
using the limits above.

Note that $\sigma_{0}, \sigma_{1}, \sigma_{2}$ do not have a definite sign, because $a_{01}\zeta,\zeta^{2}>0$,
but $a_{01}^{2}-a_{10}b_{01}<0$. Therefore we need to investigate the behaviour of this function $\sigma(\alpha)$ in more detail.

Being a quadratic function $\sigma(\alpha)$ has a single critical point (either corresponding to a maximum or a minimum) $\alpha_{crit}$. Given that $\sigma(1/3),\sigma(1)>0$, we can avoid superluminalities if and only if $1/3<\alpha_{crit}<1$ and simultaneously $\sigma(\alpha_{crit})<0$.

Computing $\frac{d\sigma(\alpha)}{d\alpha}=0$ yields for the critical point $\alpha_{crit}$
\begin{equation}
\alpha_{crit}=-\frac{\sigma_{1}}{2\sigma_{2}}=\frac{4a_{01}^{2}-3a_{10}b_{01}+16a_{01}\zeta+48\zeta^{2}}{\sigma_{2}}\,.
\end{equation}
Plugging this back into the expression for $\sigma(\alpha_{crit})$ gives the following expression
\begin{equation}
\sigma(\alpha_{crit})=4(a_{10}b_{01}-a_{01}^{2})\frac{a_{01}^{2}+8a_{01}\zeta+16\zeta^{2}-a_{10}b_{01}}{\sigma_{2}}\,.
\end{equation}
It is useful to consider
\begin{equation}
1-\alpha_{crit}=2\frac{a_{01}^{2}+4a_{01}\zeta+48\zeta^{2}-a_{10}b_{01}}{\sigma_{2}}\,.
\end{equation}
If $\alpha_{crit}<1$ then this is positive.
Similarly
\begin{equation}
\alpha_{crit}-\frac{1}{3}=\frac{1}{3}\frac{6a_{01}^{2}+24a_{01}\zeta-4a_{10}b_{01}}{\sigma_{2}}\,.
\end{equation}
If $\alpha_{crit}>1/3$ then this is positive \footnote{ When we write $num\ of\ \sigma(\alpha_{crit})$, we mean
$a_{01}^{2}+8a_{01}\zeta+16\zeta^{2}-a_{10}b_{01}$ by that, ie we are ignoring
the uninteresting factor of $4(a_{10}b_{01}-a_{01}^{2})>0$.}.

We will now show that we cannot simultaneously satisfy all the criteria
that we need to satisfy to avoid superluminalities. We consider four
cases which will exhaust all possibilities:

\paragraph{Case 1: $\sigma_{2}=0$}$\phantom{.}$\\
In this case we have
\[
\sigma(\alpha)=\sigma_{0}+\sigma_{1}\alpha
\]
Since $\sigma(1/3),\sigma(1)>0$ we know that $\sigma(\alpha)>0$
in the whole interval $1/3<\alpha<1$.

\paragraph{Case 2: $\sigma_{2}<0$}$\phantom{.}$\\
Consider $\sigma(\alpha_{crit})$. If we assume that $\sigma_{2}$
is negative, then we can avoid superluminalities if and only if the
numerator of $\sigma(\alpha_{crit})$ is positive.

However the condition that $\sigma_{2}$ is negative means that $a_{01}b_{01}>\frac{6}{5}a_{01}^{2}+\frac{24}{5}a_{01}\zeta+\frac{144}{5}\zeta^{2}$,
which implies that
\begin{equation}
num\ of\ \sigma(\alpha_{crit})=a_{01}^{2}+8a_{01}\zeta+16\zeta^{2}-a_{10}b_{01}<-\frac{1}{5}(a_{01}-8\zeta)^{2}\,.
\end{equation}
So we cannot avoid superluminalities in this case either.

\paragraph{Case 3: $\sigma_{2}>0,\zeta>0$ }$\phantom{.}$\\
Again we consider $\sigma(\alpha_{crit})$. We now assume that $\sigma_{2}$
is positive, so we need to check if numerator of $\sigma(\alpha_{crit})$
can be negative if we also assume that $\alpha_{crit}<1$, ie $a_{10}b_{01}<a_{01}^{2}+4a_{01}\zeta+48\zeta^{2}$,
and also that $\alpha_{crit}>1/3$, ie $a_{01}b_{01}<\frac{3}{2}a_{01}^{2}+6a_{01}\zeta$.

The inequality $\alpha_{crit}<1$ tells us that
\begin{equation}
num\ of\ \sigma(\alpha_{crit})=a_{01}^{2}+8a_{01}\zeta+16\zeta^{2}-a_{10}b_{01}>4a_{01}\zeta-32\zeta^{2}=4\zeta(a_{01}-8\zeta)
\end{equation}
and the inequality $\alpha_{crit}>1/3$ tells us that
\begin{equation}
num\ of\ \sigma(\alpha_{crit})=a_{01}^{2}+8a_{01}\zeta+16\zeta^{2}-a_{10}b_{01}>-\frac{1}{2}a_{01}^{2}+2a_{01}\zeta+16\zeta^{2}
\end{equation}
Now let's take $\zeta>0$. The first inequality then implies we need
$a_{01}-8\zeta<0$ to avoid superluminalities. So we set $a_{01}=8\zeta\epsilon$
for $0<\epsilon<1$ (if $\epsilon<0$ then $-a_{01}\zeta>0$). Then
the second inequality becomes
\begin{equation}
num>16\zeta^{2}(1+\epsilon-2\epsilon^{2})=16\zeta^{2}(1-\epsilon)(1+2\epsilon)>0
\end{equation}
So also in this case we are forced to have superluminalities.

\paragraph{Case 4: $\sigma_{2}>0,\zeta<0$}$\phantom{.}$\\
Now we take $\zeta<0$. The first inequality then implies we need
$a_{01}-8\zeta>0$ to avoid superluminalties. However note that both
$a_{01}$ and $\zeta$ are negative.

So we set $a_{01}=8\zeta\epsilon$ for $0<\epsilon<1$. Then the argument
is exactly the same as above, and that concludes our set of possibilities. In conclusion there is no possible way to avoid superluminalities near the origin, even if one had been so lucky as to live in a theory with specifically tuned coefficients for which the first order corrections near the origin vanished. Our result is thus generic: superluminalities are always present near the origin if the field is to be trivial at infinity and stable both at small and large distances.

\end{document}